# COMPARATIVE ANALYSIS OF NUMERICAL METHODS
# FOR PARAMETER DETERMINATION

Ivan L. Andronov,  Maria G. Tkachenko

Department "High and Applied Mathematics",
Odessa National Maritime University, Odessa, Ukraine
*tt_ari@ukr.net, masha.vodn@yandex.ua*

ABSTRACT. We made a comparative analysis of numerical methods for multidimensional optimization. The main parameter is a number of computations of the test function to reach necessary accuracy, as it is computationally "slow". For complex functions, analytic differentiation by many parameters can cause problems associated with a significant complication of the program and thus slowing its operation. For comparison, we used the methods: "brute force" (or minimization on a regular grid), Monte Carlo, steepest descent, conjugate gradients, Brent's method (golden section search), parabolic interpolation etc. The Monte-Carlo method was applied to the eclipsing binary system AM Leo.

Determination of statistically optimal parameters fitting the observations is a common task in science, particularly, in astronomy. In modeling eclipsing binary stars, there is an important problem of dependence of the optimal model parameters on each other, which may locally be described by high correlation of the deviations of these parameters near the point of optimal solution. And, extremely, by a presence of regions in multidimensional space, which correspond to nearly equal quality of the approximating light curves. There are some well-known programs based on the method of Wilson & Devinney (1971), which allow modeling of eclipsing binary stars, such as e.g. the program by Wilson (1993), Binary Maker (Bradstreet and Steelman, 2002), PHOEBE and EBAI (Prsa et al., 2012) and others. Zoła et al. (1997, 2010) presented a program for the parameter determination using the Monte-Carlo method only. Although the program works very effectively, it needs a lot of computational time, which theoretically could be decreased using other methods for faster convergence to a minimum of the test function describing "distance" between the observations and the theoretical curve.

And so we try to find the best method for the determination of the parameters of eclipsing binary stars.

For this purpose, we have used observations of one eclipsing binary system, which was analyzed by Zola et al. (2010). This star is AM Leonis, which was observed using 3 filters (B, V, R).

For the analysis, we used the computer code written by Professor Stanisław Zoła (Zoła et al. 1997, 2010). In the program, the method Monte-Carlo is implemented. As a result: the parameters were determined and the corresponding light curves (assuming statistically optimal values of the parameters) are shown in Fig. 1.

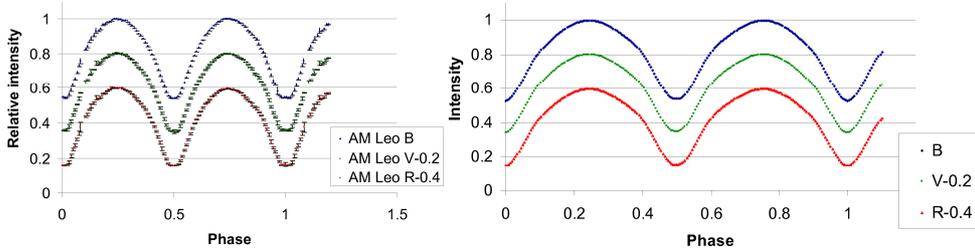

Fig. 1. A test star: AM Leo (Observations presented by S.Zoła et al. (2010)), Phase light curve of AM Leo: observations (left) and best theoretical approximation (right).

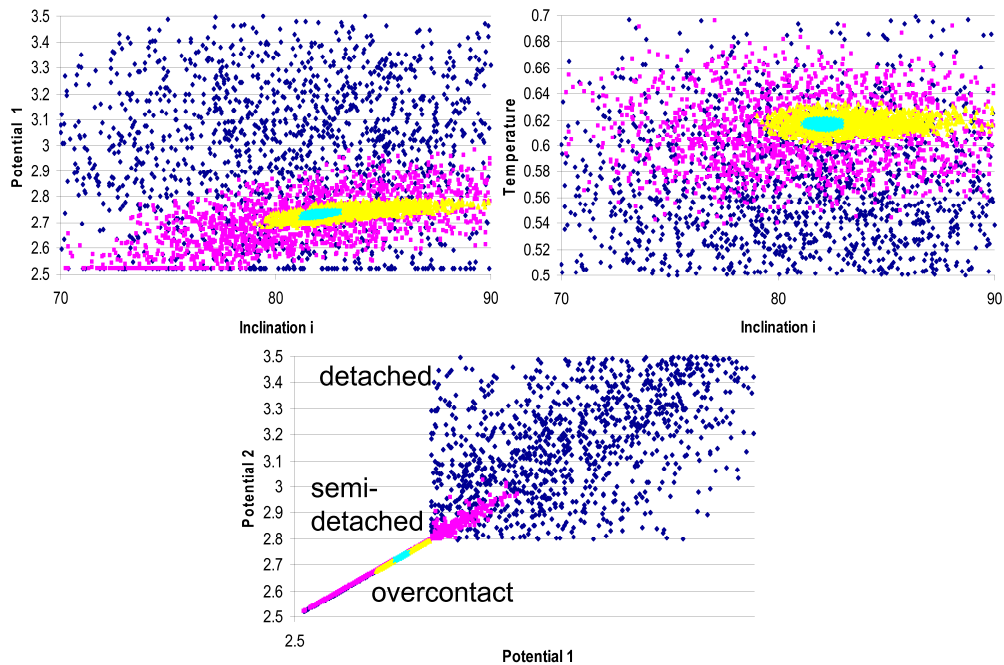

Fig. 2. Parameter–parameter diagrams: best 1500 points after hundred thousands of evaluations.

Most important pairs are: inclination versus potential of the primary or secondary star; potentials of both stars. With different colors are shown results for different numbers of computations (from which only the best 1500 points are shown). One may see that initially the points are distributed nearly homogeneously. With an increasing number of evaluations, the points are being concentrated to smaller and smaller regions. And, finally, the "cloud" should converge to a single point. Practically this process is very slow. This is why we try to find more effective algorithms. At the "potential – potential" diagram, we see that the best solution corresponds to an "over-contact" system, which makes an addition link $\Omega_1=\Omega_2$ and corresponding decrease of the number of unknown parameters.

Such a method needs a lot of computation time. We had made fitting using a hundred thousands sets of model parameters. The best 1500 (user defined) points are stored in the file and one may plot the "parameter – parameter diagrams". Of course, the number of parameters is large, so one may choose many pairs of parameters. However, some parameters are suggested to be fixed, and thus a smaller number of parameters is to be determined.

All the parameters are listed in the following table.

| Parameter | Value |
| --- | --- |
| Period $P$ | 0.44045 days |
| Eccentricity $e$ | $0^{\circ}$ (fixed) |
| Longitude of periastron $\omega$ | $90^{\circ}$ (arbitrarily) |
| Inclination $i$ | $82.2^{\circ}$ |
| Gravity darkening $g_1$ | 0.32 (fixed) |
| Gravity darkening $g_2$ | 0.32 (fixed) |
| $T_1$ | 6100 K (fixed) |
| $T_2$ | 6170 K |
| Albedo 1 | 0.5 (fixed) |
| Albedo 2 | 0.5 (fixed) |
| Potential $\Omega_1$ | 2.7397 |
| Potential $\Omega_2$ | 2.7397 |
| Mass ratio $q=M_2/M_1$ | 0.459 |
| Luminosity $L_1$ | 7.774 (arb.) |
| Luminosity $L_2$ | 4.0447 (arb.) |

Looking for the "parameter-parameter" diagrams, we see that there are strong correlations between the parameters. E.g. the temperature in our computations is fixed for one star. If not, the temperature difference is only slightly dependent on temperature, thus both temperatures may not be determined accurately from modeling. So the best solution may not be unique; it may fill some sub-space in the space of parameters.

This is a common problem: the parameter estimates are dependent. Our tests were made on another function, which is similar in behavior to a test function used for modeling of eclipsing binaries.

We have used the following test function ($x_1=x$, $x_2=y$)

$$Z(x_1,x_2)=(x_1^2-x_2)^2+\alpha\cdot(x_2-1.0201)^2. \qquad (1)$$

It is shown at Fig. 3. This function has an exact symmetric solution for a minimum for $\alpha>0$: $x_1=\pm1.01$, $x_2=1.0201$. However, if $\alpha=0$, there is an infinite number of solutions, located at the line $x_1^2=x_2$. Thus this simple function fits the criteria of complexity – two local minima instead of one global; the "ravine" is present, which links the minima; the minima are "shallow" for small $\alpha\ll1$ and completely vanishing for $\alpha=0$.

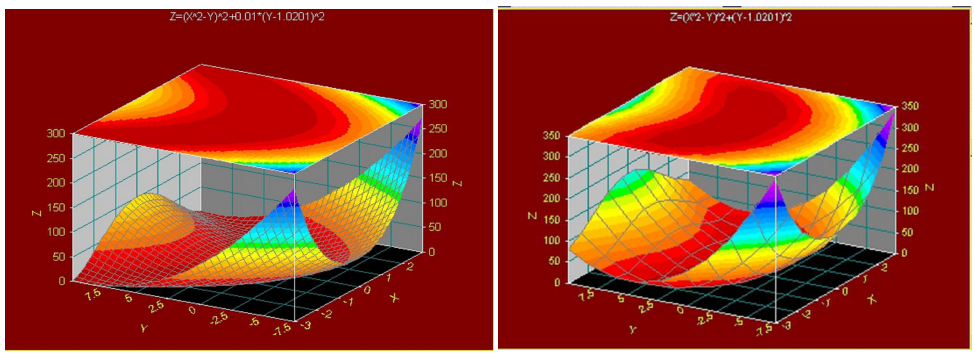

Fig. 3. 2D test function $Z(x_1,x_2)$ with 2 symmetric minima: for $\alpha=0.01$ (left) and $\alpha=1$ (right).

The tangent surface is usually defined as

$$Z(x_1,x_2)= Z(x_{01},x_{02})+ Z_1'\cdot(x_1-x_{01}) + Z_2'\cdot(x_2-x_{02})+ \qquad (2)$$
$$+0.5\cdot(Z_{11}''\cdot(x_1-x_{01})^2+2\cdot Z_{12}''\cdot(x_1-x_{01})\cdot(x_2-x_{02})+ Z_{22}''\cdot(x_2-x_{02})^2)$$

(e.g. Korn and Korn, 1968), where the derivatives $Z_j'\equiv\partial Z/\partial x_i$, $Z_{ij}''\equiv\partial^2 Z/(\partial x_i\partial x_j)$, $i,j=1..2$, are defined at the point of contact $(x_{01},x_{02})$. The shape of such surface is defined by a determinant of the Hessian matrix (combined from partial derivatives of second order). In a case of two variables, it may be defined as $D= Z_{11}\cdot Z_{22} - Z_{12}^2$.

For $D > 0$, $D = 0$ and $D < 0$, the surface is an elliptic paraboloid, parabolic cylinder and an elliptic hyperboloid, respectively.

Using Eq.(2), one may estimate a position of a stationary point (where all first derivatives vanish: $Z_j' = 0$). For $D>0$, this stationary point corresponds to an extremum (maximum for $Z_{11}<0$, else minimum), and this is a good estimate for a next point to iterate using the method of conjugate gradients. For $D<0$, the point of convergence is a "saddle" point, so it is not correct to use conjugate gradients at points with $D<0$. Instead, we suggest to use "steepest descent" and then return to "conjugate gradients", when next iteration point will correspond to $D>0$.

As expected theoretically, the fastest method should be "Conjugate gradients". However, it needs computation not only of the function, but also of derivatives of the second and first orders. At figure 5, one may see changes of the function $Z(x_1,x_2)$ with the parameter $\alpha$. At the right part, one may see 3 regions. Only in the red part, the method of the "Conjugate gradients" is working good. In this case, the graph of the function may be locally represented as an elliptic paraboloid. In the blue zone, the approximating surface is a hyperbolic paraboloid. In this case, the destination point is not a minimum, but to a singular point of the type "saddle". So we recommended to use the method of "Steepest Descent" in blue and green regions and to continue to use "Conjugate Gradients" in the red zone.

The series of images of Fig. 5 shows subsequent sets of best hundred of computations. As in binary stars, we see a sequential shrinkage of points towards parabola marked as a green line. Best points change positions with larger and larger number of computations.

We have used the search for minimal value using the Monte-Carlo method. Let $n$ be a number of computations $Z[n]=Z(x_1,x_2)$ made for randomly generated sets of parameters $(x_1,x_2)$, and $n_i$ are numbers of successive "best solutions", i.e. $Z[n_i]< Z[n_{i-1}]<\ldots< Z[n_1]$; $n_i > n_{i-1}>\ldots>n_1$. These numbers $n_i$ are very rarely distributed among all possible numbers $1..n$. Also one may compute a distance between successive "best points" $\Delta_i = n_i - n_{i-1}$.

At this plot, we see the dependence of the accuracy on the number of trial computations. One may see that this dependence is nearly linear in a double logarithmic scale. After nearly $10^9$ computations, the accuracy was close to $\delta \approx 10^{-10}$. We also computed similar dependence for other number of parameters $m$ and we see that a slope $\gamma$ at a double logarithmic scale is nearly equal to $\gamma = 2/m$. The upper line shows a dependence of a number of computations $\Delta_i$ between successes on a total number of computations $n$. The statistical dependence is linear for all values of the number of parameters $m$.

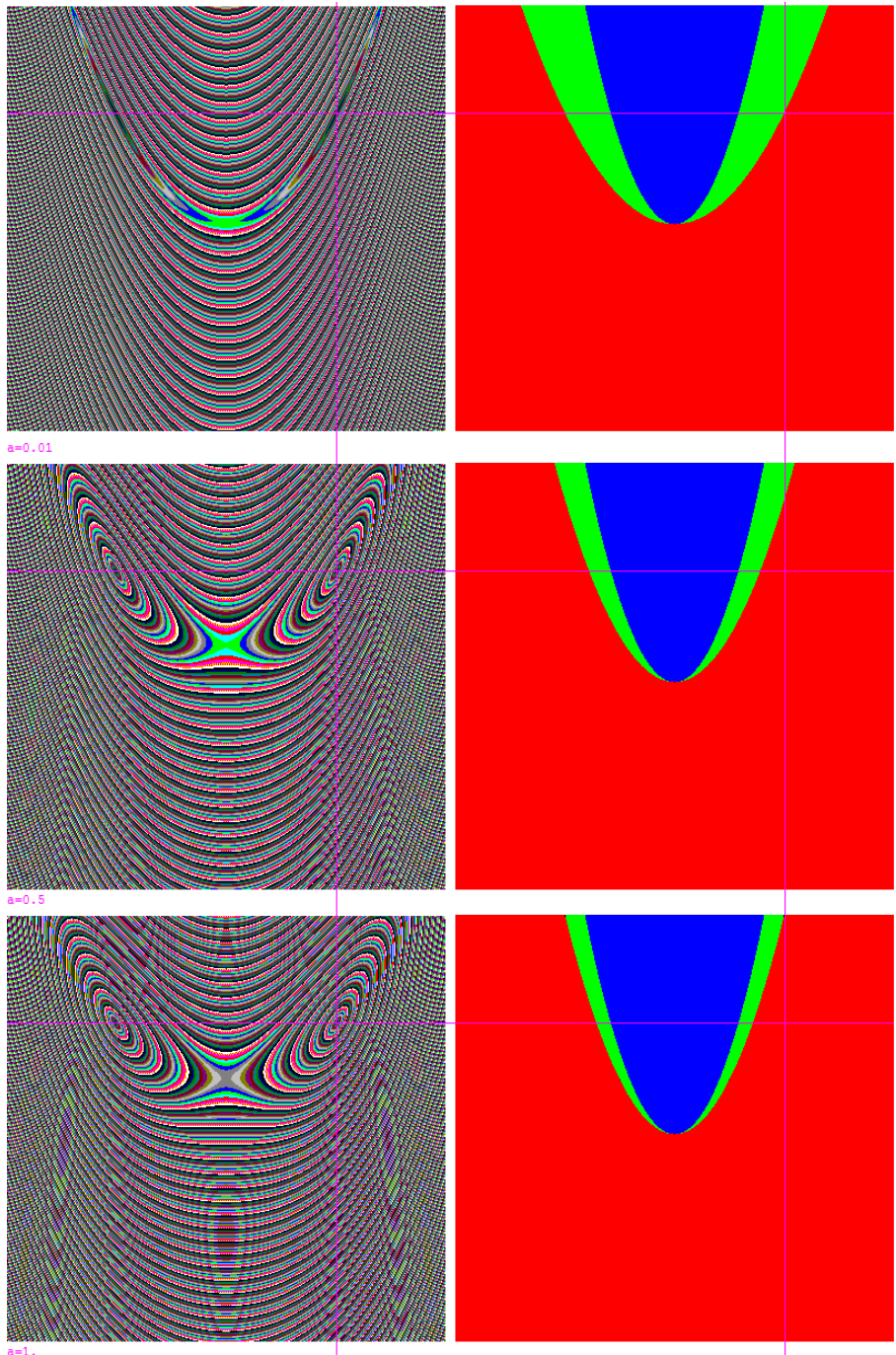

Fig. 4. Dependence of the test function $Z(x_1, x_2)$ for $-2 \leq x_1 \leq 2$, $-2 \leq x_2 \leq 2$ on the parameter $\alpha$ (0.01 (up), 0.5 (middle), 1 (bottom)). Left: levels of the function; right: zones of $D>0$, $D_{11}>0$ (red), $D<0$, $D_{11}<0$ (blue), $D<0$, $D_{11}>0$ (green).

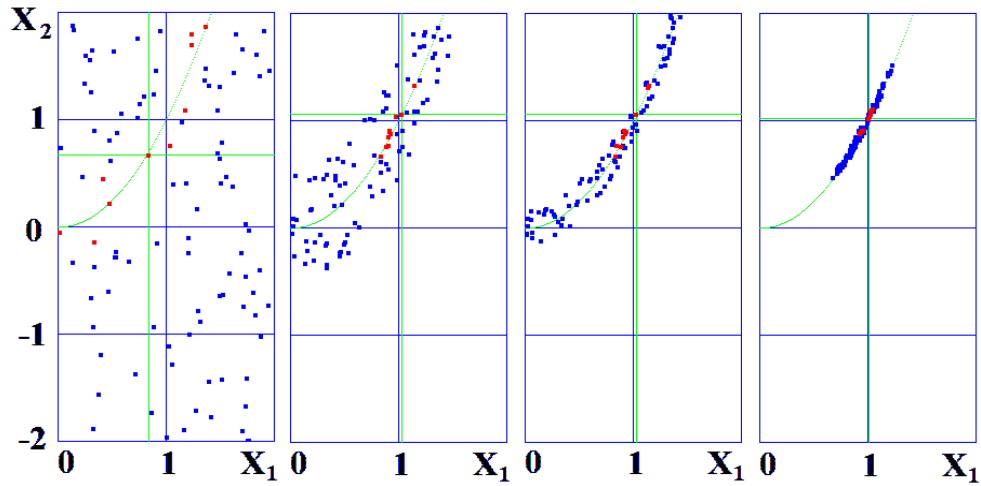

Fig. 5. Diagrams for best (100 – blue and 10 – red) pairs of points ($x_1$, $x_2$) after different number of computations (100, 600, 1600, 30000, respectively). Green cross shows the best point. Green parabola corresponds to the "ravine" $x_2 = x_1^2$.

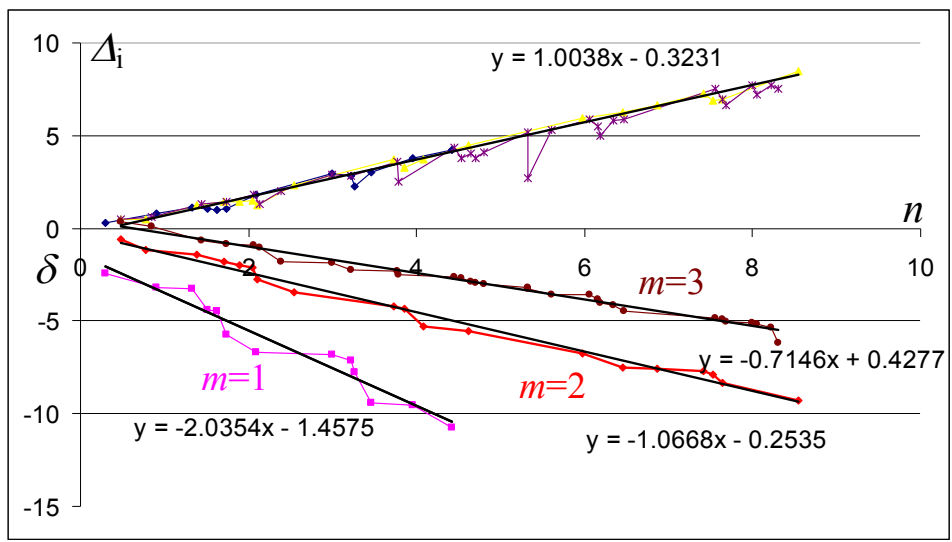

Fig. 6. Dependence of the number of trials $n$ to get deviation $\delta$ of signal from exact minimum (down) and dependence number between successful computations on a total number of computations. The graph is in a double logarithmic scale.

**We have tried also other methods:**
- Steepest descent
- Conjugate gradients
- "Brute Force" (checking on the grid)
- Minimization on subsequent directions
- Numerical estimates of derivatives of first and second order

**For 1D minimization:**
- "Brute Force" (checking on the grid with decreasing step)
- "Golden Section"
- "Dichotomy"

The corresponding programs had been written and tested.

**Future plans:**
1. To apply elaborated programs to determine parameters of spotted eclipsing binary stars.
2. To improve existing programs to a case of spots, which migrate from season to season.

*Acknowledgements.* We thank to Dr. Bogdan Wszołek and the Institute of Physics of the Jan Długosz University for hospitality and Prof. Stanisław Zoła for allowing to use the program of modeling and sample data on AM Leo. This work is a part of the projects "Inter-Longitude Astronomy" (Andronov et al., 2010) and "Ukrainian Virtual Observatory" (Vavilova et al., 2012).